# Thermal conductivity of a two-dimensional electron gas with Coulomb interaction

A. O. Lyakhov
*Department of Physics, Chernivtsi National University, 58012 Chernivtsi, Ukraine*

E. G. Mishchenko
*Bell Laboratories, Lucent Technologies, Murray Hill, NJ 07974 and
Department of Physics, University of Colorado, CB 390, CO 80309-0390*

We demonstrate that forward electron-electron scattering due to Coulomb interaction in a two-dimensional ballistic electron gas leads to the $(T \ln T)^{-1}$ temperature dependence of the thermal conductivity, which is logarithmically suppressed compared to the usual Fermi liquid behaviour.

Ballistic two-dimensional electron system with Coulomb interaction being a Fermi liquid still reveals some singularities (albeit weak) in a number of quasiparticle and collective phenomena, namely in the quasiparticle lifetime[1–9], tunneling density of states[10,11] and drag resistance[12]. In particular, at zero temperature the tunneling density of states has a cusp at the Fermi surface, $\nu(\xi) = \nu_0[1 + \frac{|\xi|}{4\epsilon_F}]$, here $\xi = \frac{p^2}{2m} - \epsilon_F$ is the distance from the Fermi energy, and $\nu_0 = m/\pi$ is the thermodynamic density of states counting both spin directions. Inverse quasiparticle lifetime contains the extra (compared to the three-dimensional case) logarithmic factor,

$$\Gamma = \frac{\xi^2}{4\pi\epsilon_F} \ln \frac{\epsilon_F}{|\xi|}. \qquad (1)$$

The numerical coefficient in this expression was subject to some confusion in the earlier works before finally being correctly established in Refs. 5,6. The logarithm in Eq. (1) is a signature of the singularity in the probability for forward, $q \simeq 0$, electron-electron scattering. Forward scattering is dominant in the high density limit, $r_s \ll 1$, here $r_s = k_0/k_F$, and $k_0 = 2\pi e^2 \nu_0$ is the inverse static screening length. This range is known to be well described by the random phase approximation (RPA). The probability of the backward scattering, $q \simeq 2k_F$, is enhanced as well. It has been shown[9] that under the assumption of low density $r_s \gg 1$, backward scattering contributes to the lifetime as much as the forward scattering does and doubles the coefficient in Eq. (1) provided that RPA is still used for the electron-electron interaction. The latter assumption however is difficult to explain since RPA becomes not a reliable approximation at $r_s \sim 1$ when correlation corrections to the 'bubble' approximation are significant. Attempts have been made to modify RPA to incorporate these corrections (see Ref. 5), but the low-density case is yet to be well understood. Moreover, the exchange effects in the electron-electron scattering as well should provide essential corrections to the 'golden rule' result (1) for $r_s \geq 1$. The importance of the exchange corrections was stressed in Refs. 7-8. One expects however only the coefficient in the lifetime to be likely to change but not the energy dependence.

It is important to investigate the effect of the electron-electron interaction singularities on the transport phenomena. Electrical resistance being related to the momentum transfer is not affected by the electron-electron interaction conserving the total momentum of scattering particles. This restriction is relaxed in a setup of two conductors spatially separated by the distance $d$, in which only the total momentum of carriers has to be conserved during a scattering event but not the momentum of carriers in each conductor. Cross-resistance appears as the result of the electron-electron interaction between carriers in different conductors, the phenomenon known as the Coulomb drag effect. It has been shown[12–14] that the singularity related to the forward interlayer scattering is suppressed by the factor $1 - \cos \phi \sim q^2/k_F^2 \ll 1$ arising in the transport cross section for the low-angle $\phi$ scattering. Therefore the drag resistance at high densities $1/d \ll k_F$ (when the backward interlayer scattering is exponentially suppressed) does not contain a singularity, $R_d \propto T^2$. Backward scattering is reported to enhance the Coulomb drag effect in the low density limit $1/d \geq k_F$ when it prevails over the forward scattering[12].

Here we consider a different transport coefficient of a 2D degenerate ballistic electron system, namely the thermal conductivity of a single 2D layer. As the thermal transport is not related to the macroscopic current the conservation of momentum by the electron-electron interaction does not impose such a restriction as in the case of the charge transport (conductivity). One can therefore expect the thermal conductivity to be reflective on features of the electron interaction. We capitalize on the high density limit $r_s < 1$ to make use of the RPA approximation where it is established to be a good one. We demonstrate that the forward scattering singularity survives and leads to the following result for the thermal conductivity,

$$\kappa = \frac{\epsilon_F^2}{\hbar T \ln \frac{\epsilon_F}{k_B T}}. \qquad (2)$$

The above argument of Refs. 12–14 about the suppression of the low-$q$ singularity does not apply in this case due to the fact that the heat transport is controlled by the factor $1 - \cos \theta$ containing the angle $\theta$ between the momenta of collided particles[15] rather that the scattering angle $\phi$. This factor does not however posess any smallness for low-$\phi$ scattering and therefore can not remove the singularity.

Now we derive this result using the Boltzmann equation for the electron-electron scattering. The thermal conductivity being the coefficient between the energy flow $\int d\mathbf{p}\, \mathbf{v}\xi_p f_p$ and the temperature gradient $\partial T/\partial x$ (applied along the $x$-axis),

$$\kappa = -\frac{1}{2}v_F^2 \nu_0 \int d\xi_p\, \xi_p \frac{\partial n_p}{\partial \xi_p}\, \psi(\xi_p), \qquad (3)$$

is determined by the nonequilibrium deviation of the distribution function $f_p$ from the Fermi-Dirac value $n_p = n(\xi_p)$,

$$f_p = n_p + \frac{\partial n_p}{\partial \xi_p}\, \chi_p, \quad \chi_p = v_F \cos\alpha_p \frac{\partial T}{\partial x}\, \psi(\xi_p) \qquad (4)$$

where $\alpha_p$ is the angle between the electron momentum $\mathbf{p}$ and the $x$-axis. The function $\psi(\xi_p) = -\psi(-\xi_p)$ is the odd function of the electron energy [more rigorously, the even part of this function $\psi_s(\xi_p)$ is nonzero and has to be found from the condition of the absence of macroscopic current[16] yielding $\psi_s \sim \psi T/\epsilon_F \ll \psi$, see also the discussion after Eq. (8)]. To find the function $\psi(\xi_p)$ we have to solve the linearized Boltzmann equation[15] which within the required approximation takes the form,

$$-v_F \cos\alpha_p \frac{\partial n_p}{\partial \xi_p}\, \xi_p \frac{\partial T}{\partial x} = 2n_p \int d\mathbf{k}d\mathbf{q}\, w(\omega,q) n_k (1-n_{p+q})(1-n_{k-q})\delta(\xi_p+\xi_k-\xi_{p+q}-\xi_{k-q})(\chi_p+\chi_k-\chi_{p+q}-\chi_{k-q}), \quad (5)$$

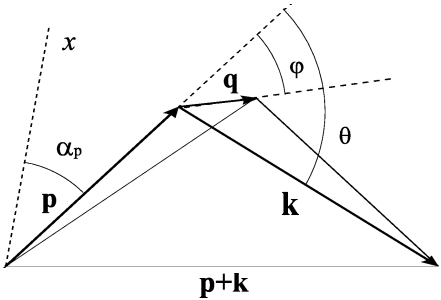

FIG. 1: Electron momenta prior $\mathbf{p}$, $\mathbf{k}$ and after $\mathbf{p}+\mathbf{q}$, $\mathbf{k}-\mathbf{q}$ a collision. The collision angle between incident momenta $\mathbf{p}$ and $\mathbf{k}$ is denoted by $\theta$, and the scattering angle (between $\mathbf{p}$ and $\mathbf{q}$) is denoted by $\phi$. The angle between $\mathbf{p}$ and the direction of the temperature gradient (the $x$-axis) is denoted by $\alpha_p$.

here $d\mathbf{k} = d^2k/(2\pi)^2$, the coefficient 2 accounts for the spin degeneracy, $w(\omega,q)$ is the probability of scattering with the transferred momentum $\mathbf{q}$ and the transferred energy $\omega = \xi_{p+q} - \xi_p$ respectively [in what follows we use the units: $\hbar = k_B = 1$]. The scattering probability in the golden rule approximation, $w(\omega,q) = 2\pi |U(\omega,q)|^2$, is determined by the matrix element of the dynamically screened Coulomb interaction which within the RPA scheme is given by $U(\omega,q) = 2\pi e^2/[q\, \varepsilon(\omega,q)]$, with the dielectric function,

$$\varepsilon(\omega,q) = 1 - \frac{4\pi e^2}{q}\int d\mathbf{p}\, \frac{n_p - n_{p+q}}{\omega + \xi_p - \xi_{p+q} + i\eta}. \qquad (6)$$

In writing Eq. (5) we have neglected the exchange effects for the scattering between electrons with the same spin direction. This assumption is justified as the dominant contribution comes from small transferred momenta $q \ll k_0$, while the exchange interaction generally involves large transfers $q \sim k_F > k_0$.

To proceed with the right-hand side of Eq. (5) we choose the energy $\xi_k$, the angle $\theta$ between vectors $\mathbf{k}$ and $\mathbf{p}$, the magnitude of the transferred momentum $q$ and the transferred energy $\omega$ [satisfying $-qv_F < \omega < qv_F$] for the new variables,

$$d\mathbf{k}d\mathbf{q} = \frac{\nu_0 q\, d\xi_k d\theta dq d\omega}{16\pi^3 \sqrt{q^2 v_F^2 - \omega^2}}.$$

The angle between vectors $\mathbf{p}$ and $\mathbf{q}$, see Fig. 1, is expressed in these variables as $\cos\phi = \omega/qv_F$. With the help of Eq. (6) we can write for the scattering probability $w(\omega,q) = 2\pi \sin^2\phi/\nu_0^2$, in the range of interest[17]. It is convenient to eliminate the delta-function by integrating first over the angle $\theta$ using the following identity, which is easy to verify,

$$\delta(\omega - \xi_k + \xi_{k-q}) = \frac{\delta(\theta - 2\varphi) + \delta(\theta)}{\sqrt{q^2 v_F^2 - \omega^2}}.$$

Note that the scattering cross section $w(\omega,q)\delta(\omega - \xi_k + \xi_{k-q})d\mathbf{k}d\mathbf{q} \propto dq/q$ has a logarihmic singularity at small transferred momenta.

The angles of scattered particles in the direction dependence of the distribution function (4) can be approximated with: $\cos\alpha_{p+q} \simeq \cos\alpha_p$, and $\cos\alpha_{k-q} \simeq \cos\alpha_k = \cos[\alpha_p + \theta]$, neglecting small corrections of order $q/k_F$. The integration over $dq$ that follows the angle integration is very simple and yields for the right-hand side of Eq. (5),

$$\cos\alpha_p \frac{\partial T}{\partial x} \frac{n(\xi_p)}{\nu_0 v_F} \int \frac{d\xi_k d\omega}{2\pi^2} \, n(\xi_k)[1-n(\xi_p+\omega)][1-n(\xi_k-\omega)] \left( [\psi(\xi_p)-\psi(\xi_p+\omega)] \ln\left[\frac{\epsilon_F^2}{\omega^2}\right] + \psi(\xi_k) - \psi(\xi_k-\omega) \right),$$

here the logarithmic integral has the upper cut-off at $q \sim k_F$. Evaluating one energy integral and using the antisymmetry property of $\psi(\xi_p)$ we obtain the integral equation for this function,

$$\frac{\epsilon_F \xi_p}{T} = \int_{-\infty}^{\infty} \frac{d\omega}{2\pi} K(\omega, \xi_p) \left[ \ln\left[\frac{\epsilon_F}{|\omega|}\right] \psi(\xi_p) - \left(\ln\left[\frac{\epsilon_F}{|\omega|}\right] + 1\right) \psi(\xi_p+\omega) \right], \qquad (7)$$

with the kernel function given by,

$$K(\omega, \xi_p) = \frac{\omega}{e^{\omega/T}-1} \frac{1-n(\xi_p+\omega)}{1-n(\xi_p)}.$$

The equatuion (7) can be solved by noting that the logarithms are slowly varying functions of the frequency $\omega$ compared to the kernel $K(\omega, \xi_p)$ and therefore may be assumed to be $\sim \ln\frac{\epsilon_F}{T}$, taken approximately as constants at the characteristic frequency of the integral equation which is simply given by the temperature $\sim T$. In addition, with the logarithmic accuracy one can disregard the unity compared to the large logarithm in the second line of Eq. (7). The resulting integral equation is of the type studied in Ref. 18 with respect to the thermal conductivity of a 3D Fermi liquid. It is solved by Fourier transforming the integral equation (7) into the second order differential equation allowing an exact solution in terms of Jacoby polinomials. We obtain,

$$\psi(\xi_p) = \frac{2\epsilon_F \cosh[\frac{\xi_p}{2T}]}{T^2 \ln\frac{\epsilon_F}{T}} \sum_{n=0}^{\infty} \frac{2n+\frac{5}{2}}{(n+\frac{1}{2})(n+2)(n+1)}$$
$$\times \int_0^{\infty} dz \, \frac{\sin[z\frac{\xi_p}{T}] \sinh[\pi z]}{\cosh^2[\pi z]} P_n^{(1,\frac{1}{2})}\left(1 - \frac{2}{\cosh^2[\pi z]}\right). \qquad (8)$$

Substituting this solution into the expression for the thermal conductivity (3) and evaluating first the integral over $d\xi_p$, then over $dz$ and finally calculating the sum over $n$ we obtain Eq. (2).

Note that despite the fact that the even part of the distribution function is small compared to its odd part, $\psi_s \sim \psi T/\epsilon_F$, making a negligible contribution to the thermal conductivity $\sim T^2/\epsilon_F^2$ [as readily seen from Eq. (3)], the nonzero value of $\psi_s$ is crucial for the steadiness of the obtained solution. The presence of only the temperature gradient makes the stationary solution of the Boltzmann equation impossible as the total momentum of electrons grows indefinitely. To get a zero macroscopic current a small compensating electric field (of the order of $\sim T\nabla T/e\epsilon_F$) has to be applied. This condition leads to another equation[16] for the even part of the distribution function $\psi_s$. Qualitatively the physical picture could be understood as follows. In a relevant experimental setup of a finite-size two-dimensional layer the electric current starts to flow upon the application of a temperature gradient. The current leads to the accumulation of the electric charges at the contacts until the required value of the compensating electric field is reached. The thermal conductivity is then determined in a standard way from the energy flow in the absence of macroscopic current[19].

The thermal conductivity of a two-dimensional electron gas with Coulomb interaction is suppressed by the large logarithmic factor compared to what might be expected for the Fermi liquid. This logarithm is a signature of a pronounced role played by the forward ($q \to 0$) electron-electron scattering. The reason for not suppressing this singularity lies in the non-elasticity nature of the electron-electron collisions. Because of that the energy flows $\frac{\mathbf{p}}{m}\xi_p f_p$ before and after a collision are not likely to cancel despite a smallness of the transferred momentum $q \ll k_F$ as the electron energies experience a significant change $\sim T$. In addition the distribution function of colliding electrons (see Eqs. (4,8)) rapidly changes at the same scale too. Therefore the incident and scattered energy flows are in general quite different, the scattering becomes 'more effective' thus suppressing heat transfer, and the thermal conductivity decreases.

The experimental observation of the temperature dependence (2) is likely to be accomplished in high quality GaAs/AlGaAs or AlGaN/GaN heterojunctions. The typical 2D electron densities $\sim 10^{12} cm^{-2}$ however correspond to the low density limit $r_s \geq 1$. Still it would be interesting to analize possible deviations from the prediction (2) which would provide further insights on the relative importance of the forward and backward scattering in 2D interacting electron systems. One should keep in mind that the phonon contribution to the thermal conductivity could easily become large (or even dominant) at low temperatures[19]. The easiest way to extract the electron contribution (2) would be to control the 2D electron density (and hence the Fermi energy) by electrostatic gating. The phonon contribution being independent of the electron density could then be separated from the strongly density-dependent electron contribution.

We are grateful to A. Andreev, I. Beloborodov and P. Silvestrov for useful discussions. E.M. acknowledges the sponsorship by the Grants DMR-9984002 and BSF-9800338 and by the A.P. Sloan and the Packard Foundations.